\documentclass{article}
\usepackage{hyperref,amssymb,amsmath,mathrsfs,bm,graphicx}
\usepackage{epsfig,graphics}
\begin{document}
\centerline{\large\bf BONDIAN FRAMES TO COUPLE MATTER WITH RADIATION}
\vspace*{0.245truein}
\centerline{\footnotesize{\large W. Barreto\footnote{Centro de F\'\i sica Fundamental,
Facultad de Ciencias, Universidad de Los Andes, M\'erida, Venezuela}
 L. Castillo\footnote{Departamento de F\'\i sica, Escuela de Ciencias, N\'ucleo de Sucre,
Universidad de Oriente, Cuman\'a, Venezuela}
and E. Barrios\footnote{Postgrado en F\'\i sica Fundamental,
Facultad de Ciencias, Universidad de Los Andes, M\'erida, Venezuela}}}
\baselineskip=12pt
\vspace*{0.21truein}
\date{\today}
\begin{abstract}
A study is presented for the non linear evolution of a self gravitating distribution of matter coupled to a massless scalar field.
The characteristic formulation for numerical relativity is used to follow the evolution
by a sequence of light cones open to the future. Bondian frames are used to endow physical meaning to the matter variables and to the massless scalar field. Asymptotic approaches to the origin and to infinity are achieved; at the boundary surface interior and exterior solutions are matched guaranteeing the Darmois--Lichnerowicz conditions.  To show how the scheme works some numerical models are discussed. We exemplify evolving
scalar waves on the following fixed backgrounds: 
A) an atmosphere between the boundary surface of an incompressible mixtured fluid and infinity; 
B) a polytropic distribution matched to a Schwarzschild exterior; C)
a Schwarzschild--Schwarzschild spacetime. 
The conservation of energy, the Newman--Penrose constant preservation and other expected features are observed.

\vspace*{0.21truein}
\end{abstract}
%\pacs{04.25.-g, 04.25.D-, 04.40.-b}
%\keywords{Characteristic Evolution; Numerical Relativity}
%\maketitle
\section{Introduction}
Numerical relativity has reached high sophistication levels to
advance in the study of realistic solutions to the Einstein equations.
\cite{seidel05}. Particularly, to simulate gravitational radiation
from collapsing sources \cite{sfmp03} and binary black hole mergers
\cite{reisswigetal09}. 
Almost all the investigations have been done in the ADM $3+1$ 
formulation \cite{adm}, although the characteristic formulation
offers a myriad of valuable advantages \cite{w05}.

This work is motivated by the possibility of simulating gravitational
radiation from axisymmetric matter sources, as part of a longer project
\cite{bglmw99}, \cite{bghlw03}. We met that purpose beginning
with the most simplified
and concomitant problem under spherical symmetry, that is, the self--gravitating
massless scalar field. In the same vein we deal with the problem
of matter coupled to radiation which eventually will lead us to get
gravitational signals from bounded sources. As far as we can see, this problem
stands alone as an important one in the field.
Here we report a study in the aforementioned direction,
limiting ourselves to spherical symmetry to follow the evolution
of a massless scalar field interacting with a perfect fluid
distribution of matter.  This model problem offers a  number
of advantageous computational and geometrical features.
 It is well known that the scalar field
mimics gravitational radiation and it has been used to study
the global properties of the spacetime, black hole threshold
and radiative signals \cite{gw92}, \cite{1}, \cite{23}, \cite{4}, \cite{5}. 

The framework for computing a complete spacetime
within the characteristic approach has been laid out by Tamburino and Winicour
\cite{tw66} and explicitly in \cite{iww83}, where is explored the production
of gravitational waves for axially symmetric distribution of matter (for a complete
review see \cite{w05} and references therein). Some years later, fundamental studies
in 1D and 3D \cite{gw92}, \cite{gomezetal94}, \cite{bishopetal96} were followed
by a 3D code able to treat matter by Bishop et al. \cite{bishopetal97}.
A relatively simpler system is a spherically
symmetric distribution of a perfect fluid coupled with scalar radiation
\cite{sfp01}. For other symmetries other than spherical, the
mathematical problem has a very similar structure \cite{sfmp03}.
In these contexts we can take full advantage of the characteristic approach
to treat matter and radiation.
The first example of the use of characteristic numerical relativity for the study of
dynamical neutron star spacetime, collapse and radiative signals was 
reported in \cite{sfp01}. Few investigations consider the scalar field 
interaction with fluid stellar distributions
\cite{hc00}, \cite{ptbs00}. 

We have recently discovered an unexpected unity in the treatment
of matter in numerical relativity \cite{b09}, using explicit Bondian observers \cite{b64}.
These observers offer an Eulerian (noncomoving) 
description (global) with the spirit of Lagrangian observers (local and comoving).
Following this line also we reported a disclosure of a central equation of
state (CEoS), which is unique for all evolutions; it emerges as a conserved quantity from the 
field equations \cite{bcb09}.

Authors commonly refer to the scalar field as a model of matter
distribution that simplifies
the treatment of the hydrodynamic issues. This approach has been
useful to study non linear physics and asymptotic behaviors, especially for central regions
\cite{g95}. Scalar field models have been extended to realistic situations such as gravitational radiation \cite{gomezetal94}.
However, cases where the scalar field is coupled with radiation
require a different approach due to both the confinement effect of matter and the dispersive
nature of radiation. When a scalar field is coupled to matter
it can be easily interpreted as an anisotropic fluid if we use Bondian frames explicitly.

We develop a  numerical framework to deal with matter coupled to scalar radiation. We perform a detailed study of the central world line at $r=0$, leading to a conformally flat spacetime in that region.
We assume that the radial dependence of the geometrical
and physical variables keep the same dependence as the static variables near the coordinate--origin.

We formulate in this work the characteristic evolution in terms of
Bondian observers \cite{b09, bcb09}. The Eulerian formulations of numerical
relativity \cite{sfp01}, \cite{nc00} actually use Bondian observers in the
mathematical treatment of matter. We also match
the interior solution with the exterior one in
a clear and precise treatment of the boundary distribution of matter without the use
of artificial atmosphere. Infinity is treated  as usual in literature.

In what follows we write the field equations for Bondian observers when matter is coupled to scalar radiation, which makes the fluid manifestly anisotropic.
Section 3 is devoted to the regularization and matching. In section 4 we show how the 
scheme works by means of numerical test models.
Finally we summarize with some remarks in section 5.
%%%%%%%%%%%%%%%%%%%%%%%%%%%%
\section{Field equations for Bondian frames}
%%%%%%%%%%%%%%%%%%%%%%%%%%%%
Bondi's metric in the spherical form reads \cite{b64}
\begin{equation}
ds^2=e^{2\beta}\Bigl(\frac{V}{r}du^2+2dudr\Bigr)
     -r^2(d\theta^2+\sin\theta^2 d\phi^2), \label{bm}
\end{equation}
where $\beta=\beta(u,r)$ and $V=V(u,r)$. In these coordinates the components of the
energy--momentum tensor are distinguished  by a bar.
In spherical symmetry
there exists a well defined notion of quasilocal energy, the Misner--Sharp mass function,
$\tilde m(u,r)$ \cite{ms64h94h96} introduced by means of
\begin{equation} 
\tilde m=\frac{1}{2}(r-Ve^{-2\beta}),\label{5} 
\end{equation}
which measures the energy content in the sphere of radius $r$ and it reduces to the
Arnowitt--Deser--Misner and Bondi masses in the appropriate limits.

Consider a stress--energy tensor
for a perfect fluid and a massless scalar field
\begin{equation}
\bar T_{\mu\nu}=\bar T^M_{\mu\nu}+\bar T^\Phi_{\mu\nu}.
\end{equation}
One can follow the Tamburino--Winicour formalism \cite{tw66}, in particular
as applied in regular spacetimes, where the foliation of light cones emanates from a freely
falling central observer \cite{iww83}--\cite{sfp01}. But following Bondi, local Minkowski coordinates $(t,x,y,z)$ are introduced by
\begin{subequations}
\begin{eqnarray}
dt&=&e^{2\beta}(1-2\tilde m/r)^{1/2}du+(1-2\tilde m/r)^{-1/2}dr,\\ 
dx&=&(1-2\tilde m/r)^{-1/2}dr,\\ 
dy&=&rd\theta,\\ 
dz&=&r\sin\theta d\phi.
\end{eqnarray}
\end{subequations}
Denoting the Minkowski components of the energy--momentum tensor by a caret we have
\begin{subequations}
\begin{eqnarray}
\bar T_{00}&=&\hat T_{00}e^{4\beta}(1-2\tilde m/r),\\
\bar T_{01}&=&(\hat T_{00}+\hat T_{01})e^{2\beta},\\
\bar T_{11}&=&(1-2\tilde m/r)^{-1}(\hat T_{00}+\hat T_{11}+2\hat T_{01}),\\
\bar T_2^2&=&\bar T_3^3=\hat T_3^3=\hat T_2^2.
\end{eqnarray}
\end{subequations}
Next one assumes that for an observer moving relative to these coordinates
with velocity $\omega$ in the radial direction, the space contains an isotropic
fluid with pressure $p$ and energy density $\rho$.

For this Bondian observer, the covariant energy--momentum tensor of matter is:
\begin{equation}
\hat T^M_{\mu\nu}=\left(
\begin{array}{cccc}
\rho & 0 & 0 & 0 \\
0 & p & 0 & 0 \\
0 & 0 & p & 0 \\
0 & 0 & 0 &  p 
\end{array}
\right).
\end{equation}
Then a Lorentz transformation readily shows that
\begin{subequations}
\begin{eqnarray}
\bar T_{00}&=&e^{4\beta}(1-2\tilde m/r)\left(\frac{\rho+p\omega^2}{1-\omega^2}\right),\\
\bar T_{01}&=&e^{2\beta}\left(\frac{\rho-\omega p}{1+\omega}\right),\\
\bar T_{11}&=&(1-2\tilde m/r)^{-1}(\rho+p)\left(\frac{1-\omega}{1+\omega}\right),\\
\bar T_2^2&=&\bar T_3^3=-p.
\end{eqnarray}
\end{subequations}
The energy--momentum tensor for the massless scalar field minimally coupled with gravity 
\begin{equation}
\bar T^\Phi_{\mu\nu}=\bar\nabla_\mu\Phi\bar\nabla_\nu\Phi -\frac{1}{2}
g_{\mu\nu}\bar\nabla^\alpha\Phi\bar\nabla_\alpha\Phi,
\end{equation}
can be read by an observer at rest in the frame of (\ref{bm}).

It can be shown that the Einstein--Klein--Gordon equations can be
written as
\begin{eqnarray}
\frac{\rho+\omega^2 p}{1-\omega^2}+\rho^\Phi + \epsilon^\Phi =
\frac{1}{4\pi r}\left(-\frac{\tilde m_{,u}e^{-2\beta}}{r-2\tilde m}
+ \frac{\tilde m_{,r}}{r}\right), \label{e1}
\end{eqnarray}

\begin{equation}
\frac{\rho-\omega p}{1+\omega} + \rho^\Phi
=\frac{\tilde m_{,r}}{4\pi r^2},
\label{e2}
\end{equation}

\begin{equation}
\frac{1-\omega}{1+\omega}(\rho+p)+\rho^\Phi+p^\Phi =
 (1-2\tilde m/r)\frac{\beta_{,r}}{2\pi r},\label{e3}
\end{equation}
\begin{eqnarray}
p+p_t^\Phi&=& -\frac{1}{4\pi}\beta_{,ur}e^{-2\beta}
+ \frac{1}{8\pi}(1-2\tilde m/r) (2\beta_{,rr} + 4\beta^2_{,r}- \beta_{,r}/r)\nonumber\\
&+& \frac{1}{8\pi r}[3\beta_{,r}(1-2\tilde m_{,r})-
\tilde m_{,rr}] \label{e4}
\end{eqnarray}
and
\begin{equation}
2(r\Phi)_{,ur}=r^{-1}[r e^{2\beta}(r-2\tilde m)\Phi_{,r}]_{,r}, \label{we1}
\end{equation}
where the comma denotes partial differentiation respect to the indicated coordinate, and
the scalar energy flux $\epsilon^\Phi$, the scalar energy density $\rho^\Phi$, the scalar radial pressure $p^\Phi$,
the scalar tangential pressure $p_t^\Phi$, respectively are defined by
\begin{equation}\epsilon^\Phi=e^{-2\beta}[e^{-2\beta}(1-2\tilde m/r)^{-1}\Phi_{,u}^2 -\Phi_{,u}\Phi_{,r}],
\label{rad_phi}\end{equation}
\begin{equation}\rho^\Phi = p^\Phi = (1-2\tilde m/r)\Phi_{,r}^2/2,\end{equation}
\begin{equation} p_t^\Phi=\Phi_{,u}\Phi_{,r}e^{-2\beta}-p^\Phi.\label{tan_phi}\end{equation}
From this point of view the scalar field can be interpreted as a
radiating and anisotropic fluid \cite{hs97} whose energy--momentum tensor can be 
written as
\begin{equation}
T^\Phi_{\mu \nu}=(\rho^\Phi+p_t^\Phi)u_{\mu}u_{\nu}+\epsilon^\Phi
 l_{\mu}l_{\nu}-p_t^\Phi g_{\mu \nu}+(p^\Phi-p_t^\Phi)
\chi_{\mu}\chi_{\nu},
\end{equation}
with $u^\mu u_\mu=1,$
$l^{\mu}l_{\mu}=0,$
$\chi^\mu \chi_\mu=-1,$
if we identify the four velocity for an observer at rest in the frame of (\ref{bm}), 
the null and the space--like vectors as
\begin{equation}
u^\mu=(1-2\tilde m/r)^{-1/2}e^{-2\beta}\delta^\mu_0, \label{4v}
\end{equation}
\begin{equation} l_\mu=(1-2\tilde m/r)^{1/2}e^{2\beta}\delta^0_\mu,\end{equation}
\begin{equation}\chi_\mu=(1-2\tilde m/r)^{-1/2}\delta^1_\mu.\end{equation}
Note that Bondian observers can be purely Lagrangian when we deal only with radiation \cite{b09}. 

%........................
The conservation equation $T^\mu_{1;\mu}=0$, or
equations (\ref{e2})--(\ref{e4}), lead 
us to the generalized Tolman--Oppenheimer--Volkoff (TOV) equation for non static radiative
situations
\begin{eqnarray}
&&\tilde p_{,r}-e^{-2\beta}\Bigl(\frac{\tilde{\rho}+\tilde{p}}
{1-2\tilde{m}/r}\Bigr)_{,u}
+\Bigl(\frac{\tilde{\rho}+\tilde{p}}{1-2\tilde{m}/r}\Bigr)
\Bigl[4\pi r (\tilde{p}+p^\Phi_r)+\tilde{m}/r^{2}\Bigr]\nonumber\\
&=&\frac{2}{r}(p-\tilde{p}),\label{tov}
\end{eqnarray}
where
\begin{equation}
\tilde{\rho}=\frac{\rho-\omega p}{1+\omega},
\end{equation}
\begin{equation}
\tilde{p}=\frac{p-\omega\rho}{1+\omega},
\end{equation}
are the named effective variables \cite{hjr80},

From the field equation (\ref{e1}) is straightforward that
\begin{equation}
\frac{d\tilde m}{du}=-4\pi r^2[(p-p^\Phi)\frac{dr}{du}+
  (1-2\tilde m/r)e^{2\beta}\epsilon^\Phi], \label{mdu}
\end{equation}
where
\begin{equation}
\frac{dr}{du}=e^{2\beta}(1-2\tilde m/r)\frac{\omega}{1-\omega}, \label{mv}
\end{equation}
is the matter velocity.  Integrating the wave equation (\ref{we1}) we obtain
\begin{equation}
2r\Phi_{,u}=e^{2\beta}(r-2\tilde m)\Phi_{,r}+
\int^r_0 e^{2\beta}(1-2\tilde m/r)\Phi_{,r}dr,
\end{equation}
which combined with (\ref{mdu}) lead us to
\begin{eqnarray}
\frac{d\tilde m}{du}&=&-4\pi e^{2\beta}N^2-4\pi r^2\frac{dr}{du}p\nonumber\\
&+& 2\pi r^2(1-2\tilde m/r)\left(\frac{1+\omega}{1-\omega}\right)e^{2\beta}\rho^\Phi, \label{mdu2}
\end{eqnarray}
where
$$N=\frac{1}{2}e^{-2\beta}\int^{r}_{0} e^{2\beta}(1-2\tilde m/r)\Phi_{,r}dr.$$
In absence of matter, $\rho=p=0$, for the exterior region the field
equations reduce to 
\begin{equation}
R_{\mu\nu}=-8\pi\Phi_{,\mu}\Phi_{,\nu},
\end{equation}
or explicitly to the hypersurface equations
\begin{equation}
\beta_{,r}=2\pi r \Phi_{,r}^2, \label{ekg+1}
\end{equation}
\begin{equation}
\tilde m_{,r}= 2\pi r (r-2\tilde m) \Phi_{,r}^2, \label{ekg+2}
\end{equation}
and to the wave equation (\ref{we1}) \cite{gw92}.

%%%%%%%%%%%%%%%%%%%%%%%%%%%%%%%%%%
\section{Regularization and matching}
%%%%%%%%%%%%%%%%%%%%%%%%%%%%%%%%%%
Some previous investigations consider regularization near $r=0$ \cite{sfp01}, \cite{nc00}, \cite{lake}, \cite{bona-masso}, \cite{alcubierre}. The conditions
for the scalar field, as a matter model, do not necessarily apply to distributions of matter. 
Depending on gauge conditions each procedure to get regular spacetimes may be
cumbersome and tricky, even in vacuum. Initially regular spacetimes can eventually develop singularities \cite{g95}, \cite{choptuik}.
We show a simple way to construct regular spacetimes, near the coordinate--origin, when
the inner spacetime corresponds to a spherical distribution of baryonic matter coupled
to a massless scalar field.
To construct regular and general enough spacetimes, which eventually
recover equilibria, collapse, form singularities and horizons, 
we do an asymptotic study close to the special regions: $r=0$ and $r\rightarrow\infty$. 
The treatment is basically the same for these two zones, that is, power expansions of $r$ and $r^{-1}$, respectively. For $r=R(u)$, the boundary surface, the 
Darmois--Lichnerowicz \cite{dl},\cite{hj83} conditions are guaranteed to match the interior and exterior solutions on a moving boundary. Integrating from $r=0$ no additional conditions are required at the surface to describe its evolution.
%%%%%%%%%%%%%%%%%%%%%%%%%
\subsection{Close to the origin}
%%%%%%%%%%%%%%%%%%%%%%%%%
Consider the following asymptotic expansions for the metric functions near $r=0$,
which represents a regularly and conformally flat spacetime \cite{bcb09}:
\begin{equation}
\tilde m=m_3(u) r^3 + \mathcal{O}(r^5), \label{mto0}
\end{equation}
\begin{equation}
\beta=\beta_0(u)+ \beta_2(u) r^2 +  \mathcal{O}(r^4), \label{betato0}
\end{equation}
and for the escalar field
\begin{equation}
\Phi=\Phi_0(u) + \Phi_1(u) r + \Phi_2(u) r^2  + \mathcal{O}(r^3). \label{Phito0}
\end{equation}
Using the field equations we get the physical variables expansion as a function of $r$ as showed in Table I. 
\begin{table}[!ht]
\begin{center}
\begin{tabular}{|l|c|c|c|}
\hline
$\mathcal{O}$ &$\pi\rho$&$\pi p$&$\omega$\\
\hline
$0$&$\frac{3}{4}m_3-\frac{1}{2}\pi\Phi^2_1$&$\beta_2-\frac{3}{4}m_3-\frac{1}{2}\pi\Phi^2_1$&$0$\\
\hline
$1$  & $\omega_1(\beta_2-\pi\Phi^2_1)+\pi\Phi_1\Phi_2$  & $\omega_1(\beta_2-\pi\Phi^2_1)-2\pi\Phi_1\Phi_2$  &  $\omega_1$        \\
\hline
\end{tabular}
\caption{Coefficients for the physical variables expansion as a power of $r$ near the center.}
\label{tab:expansion}
\end{center}
\end{table}

From the evolution equation (\ref{we1}) we get
\begin{equation}
e^{-2\beta_0}\frac{ d\Phi_0}{du} = \Phi_1, \label{dotPhi0}
\end{equation}
and
\begin{equation}
 e^{-2\beta_0}\frac{d\Phi_1}{du} = \frac{3}{2}\Phi_2. \label{dotPhi1}
\end{equation}
From the field equation (\ref{e4})
\begin{equation}
 e^{-2\beta_0}\frac{d\beta_2}{du}=[\pi\Phi_1\Phi_2-2\omega_1(\beta_2-\pi\Phi^2_1)],  \label{dotbeta}
\end{equation}
and from (\ref{e1})
\begin{equation}
\frac{dm_3}{du}=2\frac{d\beta_2}{du}\label{dotem}.
\end{equation} 
This last equation, together with expansions showed in Table I, 
is readily integrated to give
\begin{equation}
\rho_c+3p_c =\mbox{constant} \label{ee_0},
\end{equation}
where $\rho_c=\rho_0+\rho_0^\Phi$  and $p_c=p_0+p_0^\Phi$.

The lapse of the coordinate time $du$ is related to the corresponding lapse of time 
$d\tau$ measured by a central observer as
\begin{equation}
d\tau=e^{2\beta_0}du.\label{propertime}
\end{equation}
We have preference for the central time $\tau$ to describe the studied system. The reason is
mainly numerical: proceeding with the radial integration from $r=0$, where the world line
is therefore geodesic, in consistence with the conformally flat result as a consequence of regularity.
Therefore, the replacements $\beta\rightarrow\beta-\beta_0$ and $u\rightarrow\tau$ left
invariant the field equations (\ref{e1})--(\ref{we1}).
The same situation will be analog for the matching surface, which behaves
asymptotically as Vaidya or Schwarzschild, and for the asymptotically flat infinity. 

%%%%%%%%%%%%%%%%%%%%% 
\subsection{Matching at the surface}
%%%%%%%%%%%%%%%%%%%%%
Boundary conditions at the surface $r=R(u)$ are needed for $\beta$ and $\tilde m$
in order to perform radial integrations.
We match the interior solution with the exterior at $r=R$
by means of the Darmois--Lichnerowicz conditions. These conditions
are equivalent to the continuity of the functions $\beta$ and $\tilde m$
across the boundary, and to the continuity of the spin coefficient \cite{hj83}
\begin{equation}
\gamma=(1-2\tilde m/r)\beta_{,r}-\frac{\tilde m_{,r}}{2r} - \beta_{,u}e^{-2\beta}.
\end{equation}
Considering the expansion of $\beta$ around the surface, we have
\begin{equation}
\beta_{,u}^\pm=\frac{dB}{du} - \frac{dR}{du}\beta_{,r}^\pm,
\end{equation}
where the superscript $\pm$ indicates the evaluation of the function at $r=R+0$ or
$r=R-0$, and $B=\beta(u,r=R)$. Therefore,  $\gamma^+=\gamma^-$ leads us to
\begin{equation}
\omega_R=1-2RF\frac{\delta\beta_{,r}}{\delta\tilde m_{,r}},
\end{equation}
where $F=1-2M/R$, $M=\tilde m_R$, the subscript $R$ indicates that the quantity is evaluated at the surface, and $\delta\Psi=\Psi^+-\Psi^-$ represents the jump
of the indicated function across the boundary.
Observe that $\delta\dot R=0$, where over dot indicates derivative respect to time.
Thus, we get 
\begin{equation}
\tilde p_R+\omega_R\tilde\rho_R=(1+\omega_R)\delta\rho^\Phi,
\end{equation}
which leads us directly to
\begin{equation}
p_R=\left(\frac{1+\omega_R}{1-\omega_R}\right)\delta\rho^\Phi. 
\end{equation} 
If the scalar field gradient is continuous across the boundary $r=R$ we have a pressureless surface, which can be expressed as $\tilde p_R=-\omega_R\tilde \rho_R$ or equivalently
\begin{equation}
\omega_R=\frac{\tilde m_{,r}^- -2RF\beta_{,r}^- + 4\pi R^2\rho^\Phi_R}
{\tilde m_{,r}^- -4\pi R^2\rho^\Phi_R}. \label{omegaR}
\end{equation} 
When $\rho_R=0$, we solve as usual the indetermination taking the limit
\begin{equation}
\omega_R=-\lim_{r\rightarrow R}\frac{\tilde p}{\tilde \rho}=
-\frac{\tilde p_{,r}}{\tilde\rho_{,r}}\Big|_R, \label{omRII}
\end{equation} 
where
\begin{eqnarray}
\tilde p_{,r}&=& \frac{2}{r}(\tilde\rho+\rho^\Phi)  -\frac{\tilde m_{,rr}}{4\pi r^2}+ \frac{\beta_{,rr}}{2\pi r^2}(r-2\tilde m)\nonumber\\
&+& \frac{\beta_{,r}}{2\pi r^2}(4\tilde m - 2\tilde m_{,r} -1) - \rho^\Phi_{,r}
\end{eqnarray}
and
\begin{equation}
\tilde \rho_{,r}=\frac{\tilde m_{,rr}}{4\pi r^2}- \frac{2}{r}(\tilde\rho+\rho^\Phi) - \rho^\Phi_{,r}.
\end{equation}
Two comments are in order here. First,  
(\ref{omRII}) can be obtained from the continuity of the spin coefficient gradient $\gamma_{,r}^+=\gamma_{,r}^-$. Second, exactly the same result is obtained from the field equations, that is, equation (\ref{omRII}) proceed from (\ref{tov}) evaluated at the surface \cite{hjr80}. 

Once satisfied the matching conditions across the boundary $r=R$ we need
surface equations to follow the radius evolution and the exchange of energy
on it. Evaluating (\ref{mdu2}) at the surface and defining 
$$\Omega=(1-\omega_R)^{-1}$$
and 
$${\cal R}^\Phi=4\pi R^2\rho^\Phi_R,$$
we get
\begin{equation}
e^{-2B}\frac{dM}{du}=-4\pi N_{R}^{2}+\frac{1}{2}{\cal R}^\Phi(2\Omega-1)F, \label{Mdu}
\end{equation}
which clearly establishes the transfer of energy at the boundary of the matter distribution. 
Additionally, evaluating (\ref{mv}) at the surface we get
\begin{equation}
e^{-2B}\frac{dR}{du}=F(\Omega-1), \label{adu}
\end{equation}
which conforms together with (\ref{omegaR}) and (\ref{Mdu})  the system of equations
at the surface for $\rho_R\ne 0$. These equations determine completely the evolution at the surface distribution.

%%%%%%%%%%%%%%%%%%
\subsection{Close to infinity}
%%%%%%%%%%%%%%%%%%
This section is standard in literature but we include here a r\'esum\'e for the sake
of completeness. 
Assuming that the scalar field has an asymptotic expansion \cite{gw92}
\begin{equation}
\Phi=\frac{Q_1(u)}{r}+ \frac{Q_2}{r^2} + \mathcal{O}(r^{-3}),
\end{equation}
the metric functions read
\begin{equation}
\beta=H(u) - \frac{\pi Q_1^2}{r^2} + \mathcal{O}(r^{-3}),
\end{equation}
\begin{equation}
\tilde m = {\cal M}(u) - \frac{2 \pi Q_1^2}{r}
+ \frac{2 \pi Q_1 ({\cal M} Q_1-2Q_2)}{r^2} + O(r^{-3}).
\end{equation}
The coefficient $Q_1$ is the scalar monopole moment and  $Q_2$ the Newman--Penrose constant. The asymptotic Bondi mass ${\cal M}$ also can be expressed globally as
\begin{equation}
{\cal M}(u)= 4\pi\int^R_0 r^2 \tilde \rho dr + 4\pi\int^\infty_0 r^2 \rho^\Phi dr=\tilde m|_{{\cal J}^+},\label{bondi_mass}
\end{equation}
and the scalar news function \cite{gw92}
\begin{equation}
{\cal N}(u)=\frac 1 2 e^{-2H}\int^\infty_0 e^{2\beta}(1-2\tilde m/r)\Phi_{,r}dr. \label{news}
\end{equation}
The Bondi mass loss equation is
\begin{equation}
e^{-2H}\frac{d{\cal M}}{du}=-4\pi{\cal N}^2. \label{Bdu}
\end{equation}

With these definitions, the total radiated energy can be established

\begin{equation}
\Delta\equiv{\cal M}(u) - {\cal M}(u_0)= -J(u) \label{BM},
\end{equation}
where
\begin{equation}
J=4\pi\int^{u}_{u_0}{\cal N}^2 e^{2H} du.\label{Jota}
\end{equation}
It is easy to check from (\ref{bondi_mass}) that the most general Killing propagator to get
the energy conservation in the spherical context is 
\begin{equation}\xi^\mu=e^{-2\beta}\delta^\mu_0. \label{Killing}\end{equation}
As a matter of fact, we get (\ref{Killing}) when the Linkage \cite{tw66}, a finite representation of the Bondi--Metzner--Sachs asymptotic group, 
\begin{equation}
{\cal{C}}=\int T_{\mu}^{\nu} \xi^\mu d\Sigma_\nu,
\end{equation}
is compared with (\ref{bondi_mass}).

Up to now the system has been described with enough generality to proceed
with the hydrodynamical solver developing, which was not planted as a goal in this
paper. However to show how the scheme works some
numerical models are discussed.

%%%%%%%%%%%%%%%%%%%%%%%%%%%%%%%%%%%
\section{Models}
%%%%%%%%%%%%%%%%%%%%%%%%%%%%%%%%%%%
%The examples showed in what follows should be considered as toy models.  

%%%%%%%%%%%%%%%%%
\subsection{Ghost scalar fluid}
%%%%%%%%%%%%%%%%%
This model exemplifies the junction conditions, the scalar field 
energetics and the preservation of the Newman--Penrose constant. 

Consider a mixture of two components fluid of pressure
 $p+p^\Phi\rightarrow p$ and energy density $\rho+\rho^\Phi\rightarrow\rho$ which is incompressible and remains static if a layout outer scalar field guarantee the Darmois--Lichnerowiczs conditions.
Clearly from (\ref{omegaR}) these conditions  reduce to 
\begin{equation}
\tilde m_{,r}^- -2RF\beta_{,r}^- + 4\pi R^2\rho^\Phi_R=0,
\end{equation}
which is satisfied if
\begin{equation}
F\beta_{,r}^+=2\pi R \rho_R.
\end{equation}
Under this scenario $\omega=0$ everywhere, and derivatives respect to the timelike coordinate dropping to zero. The set of equations (\ref{e1})--(\ref{e4}) simplify considerably to get 
the well known interior solution \cite{hjr80}
\begin{equation}
\tilde m=M(r/R)^3,
\end{equation}
\begin{equation}
e^{2\beta} = \frac{1}{2}\left\{3\left(\frac{F}{\zeta}\right)^{1/2}-1\right\},
\end{equation}
\begin{equation}
p=\rho\left\{\frac{F^{1/2}-\zeta^{1/2}}{\zeta^{1/2}-3 F^{1/2}}\right\},
\end{equation}
where 
\begin{equation}
\zeta=[1-(1-F)(r/R)^2]
\end{equation}
and
\begin{equation}
\rho=\frac{3(1-F)}{8\pi R^2}=\frac{3M}{4\pi R^3}.
\end{equation}
Observe that $\beta_R=0$, $F\beta_{,r}^-=4\pi R\rho_R=3M/R^2$ and $p_R=0$. In this way we have constructed a ``fixed'' background of a incompressible perfect fluid coupled to an exterior scalar field which
remains ``frozen" in some hydrodynamic characteristic time scale.

\begin{table}[!ht]
\begin{center}
\begin{tabular}{|l|c|c|}
\hline
$\Phi$ & Family & Parameters  \\
\hline
$\lambda/(R+r)$  &  (a)  & $\lambda$, R \\
\hline
$\lambda(r_a-r)^4(r_b-r)^4/[(r_b-r_a)/2]^8$  &  (b) & $\lambda$, $r_a$, $r_b$  \\
\hline
$\lambda \exp{(r-r_0)^2/\sigma^2}$  & (c)  & $\lambda$, $\sigma$, $r_0$ \\
\hline
\end{tabular}
\caption{Exterior/Initial data set for the massless scalar field}
\label{tab:datasets}
\end{center}
\end{table}
\begin{figure}[!ht]
\begin{center}\includegraphics[width=4in]{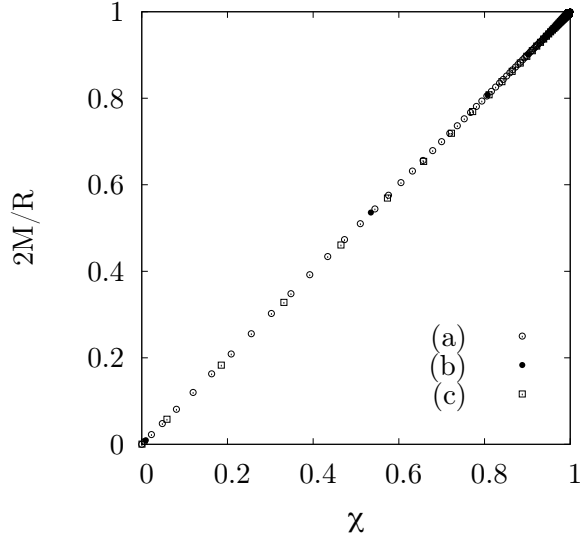}\end{center}
\caption{Mass as a function of the normalized amplitude $\chi=\pi\lambda^2/(6R^2+\pi\lambda^2)$, for the exterior datum (a).
If the datum is (b) the amplitude is $\lambda \rightarrow 13.46\lambda$; if the datum is (c) 
$\lambda\rightarrow\pi\lambda$. For any choice of the initial datum and any choice of parameters
$r_a$, $r_b$, $\sigma$ and $r_0$ --which leave partially immersed the scalar field in the fluid distribution-- the overlaping is true finding the appropriate rescaling in the amplitude.}
\label{fig:mass}
\end{figure}
\begin{figure}[!ht]
\begin{center}\includegraphics[width=4in]{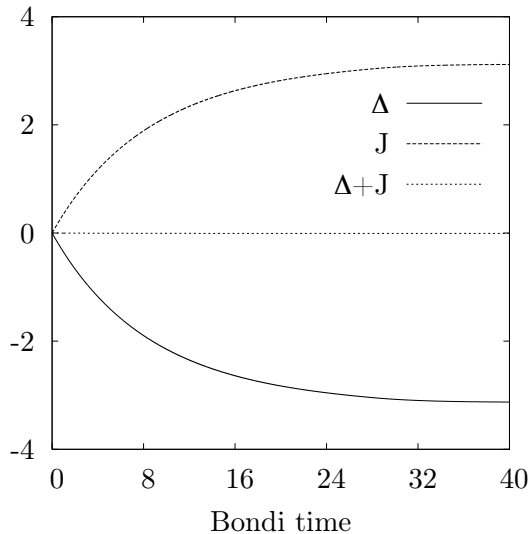}\end{center}
\caption{Energy conservation (multiplied by $10^5$) as a function of
the Bondi time for the exterior/initial datum (a). The parameters and conditions of integration are: $M=1.6921$, $R=5$, $\lambda=10$ and $10^3$ radial points. 
The evolution corresponds to a relativistic
case with $F=0.3232$. The descending curve corresponds to $\Delta$ with $\cal{M}$ calculated
by Eq. (\ref{bondi_mass}). The ascending curve corresponds to the energy radiated to infinity 
given by Eq. (\ref{Jota}) with $\cal{N}$ calculated by means of Eq. (\ref{news}). Thus, in accordance with Eq. (\ref{BM}), the horizontal curve represents the global conservation
of energy.}
\label{fig:balance}
\end{figure}
\begin{figure}[!ht]
\begin{center}\includegraphics[width=4in]{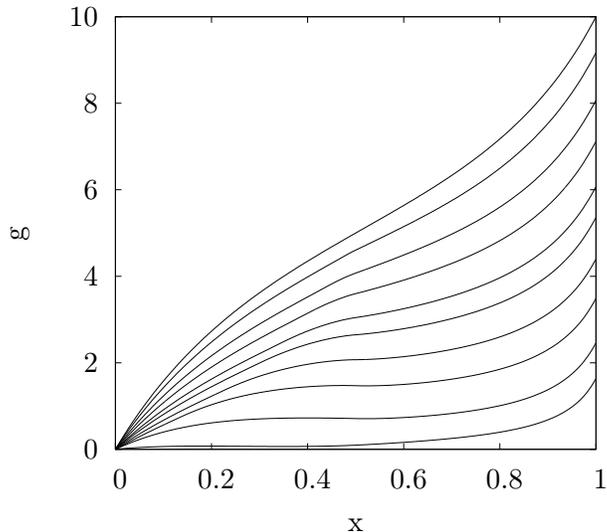}\end{center}
\caption{Decay of the scalar field (multiplied by $10^3$) preserving the Newman--Penrose constant, that is,
$g_{,x}|_{\cal{J}^+}=$ constant, any time of the evolution. The compactified coordinate $x$ is related with $r$ by means of $r=15Rx/[8(1-x^4)]$. The initial datum, the parameters and conditions are the same as for figure \ref{fig:balance}}.
\label{fig:npc}
\end{figure}
\begin{figure}[!ht]
\begin{center}\includegraphics[width=4in]{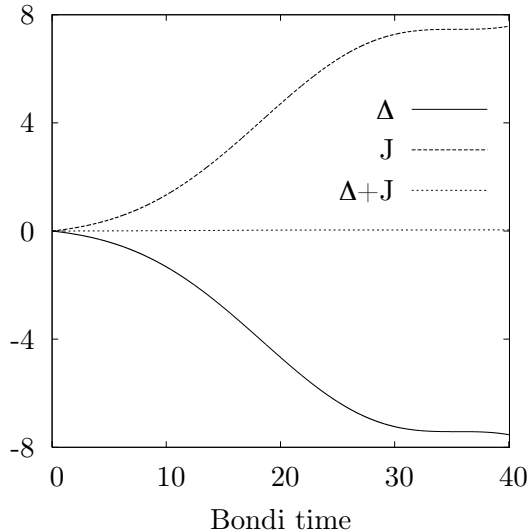}\end{center}
\caption{Energy conservation (multiplied by $10^3$) as a function of
the Bondi time for the initial datum (c). In this case the matter fluid corresponds to a polytrope with $\rho_0=4\times 10^{-3}$, $K=100$ and $\Gamma=2$.
The parameters and conditions of integration are: $M=0.774$, $R=11.512$, $\lambda=10^{-2}$,
$\sigma=R$, $r_0=R$ and $10^3$ radial points; the evolution corresponds to a $F=0.865$. The descending curve corresponds to $\Delta$ with $\cal{M}$ calculated
by Eq. (\ref{bondi_mass}). The ascending curve corresponds to the energy radiated to infinity 
given by Eq. (\ref{Jota}) with $\cal{N}$ calculated by means of Eq. (\ref{news}). Thus, in accordance with Eq. (\ref{BM}), the horizontal curve represents the global conservation
of energy.}
\label{fig:poly_cons}
\end{figure}
\begin{figure}[!ht]
\begin{center}\includegraphics[width=4in]{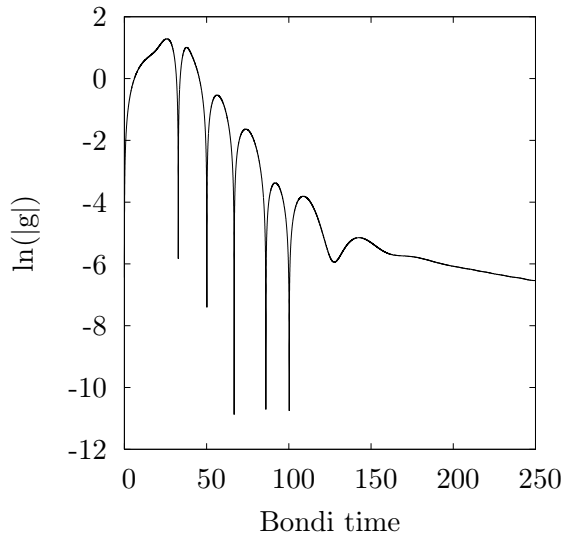}\end{center}
\caption{Quasinormal mode and tail decaying as a function of
the Bondi time for the initial datum (b). In this case the matter distribution corresponds to an incompressible fluid, that is, a Schwarzschild interior; the exterior is a Schwarzschild vacuum.
The parameters and conditions of integration are: $M=1$, $R=2.857$, $\lambda=1$, $r_a=0$, $r_b=10$, and $10^3$ radial points; the evolution corresponds to a $F=0.3$. The late time behavior of the signal decays as an inverse power law. In the window of time $[200,245]$ the power of the tail decay is approximately $-2.1$, which corresponds to a $5\%$ deviation from the expected value of $-2$ \cite{gws94}, \cite{23}.}
\label{fig:qnm}
\end{figure}
\begin{figure}[!ht]
\begin{center}\includegraphics[width=4in]{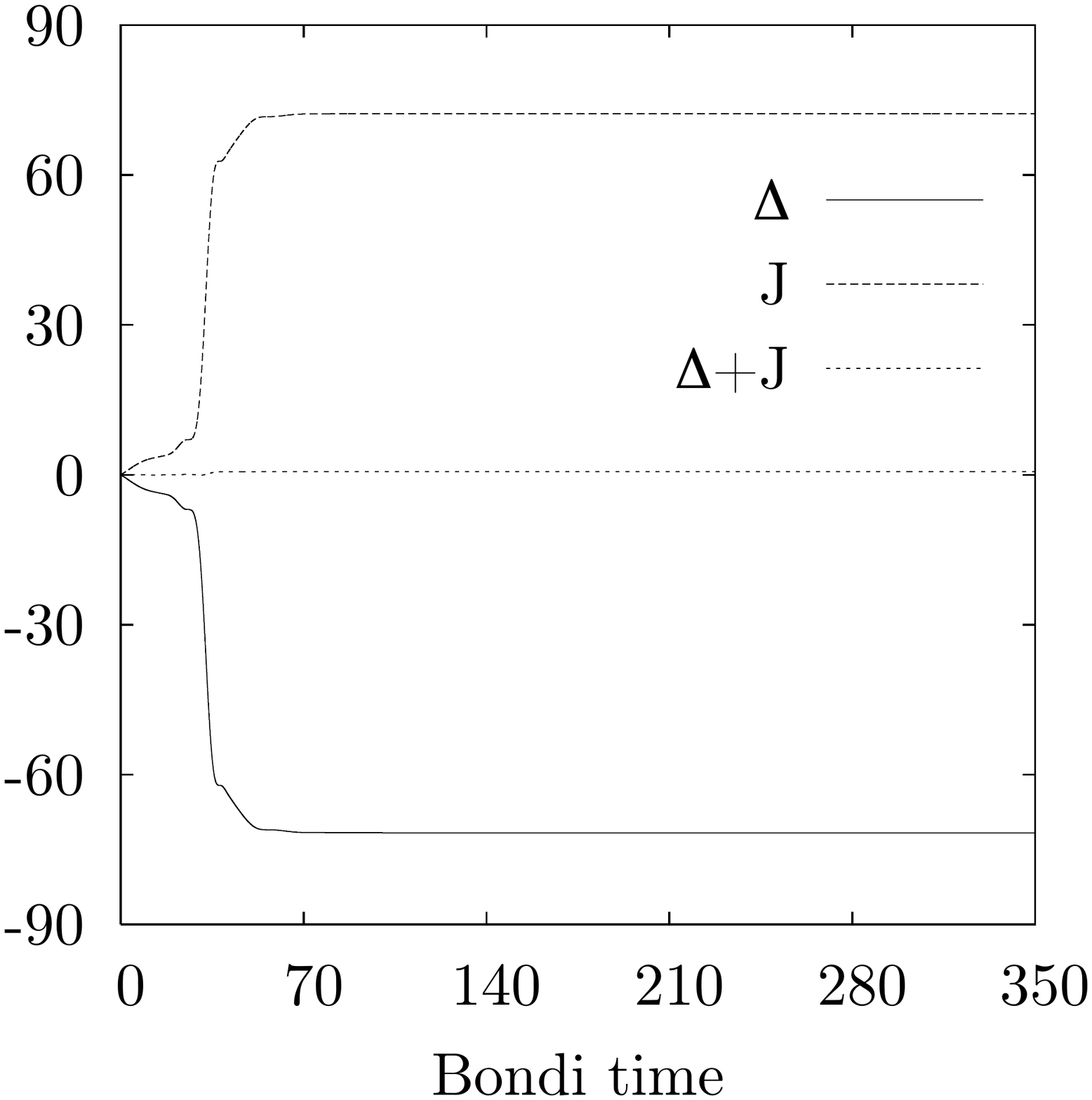}\end{center}
\caption{Energy conservation as a function of
the Bondi time for the initial datum (b). In this case the matter distribution corresponds to an incompressible fluid, that is, a Schwarzschild interior; the exterior is a Schwarzschild vacuum.
The parameters and conditions of integration are: $M=1$, $R=2.857$, $\lambda=1$, $r_a=0$, $r_b=10$, and $10^3$ radial points; the evolution corresponds to a $F=0.3$. The descending curve corresponds to $\Delta$ with $\cal{M}$ calculated
by Eq. (\ref{bondi_mass}). The ascending curve corresponds to the energy radiated to infinity 
given by Eq. (\ref{Jota}) with $\cal{N}$ calculated by means of Eq. (\ref{news}). Thus, in accordance with Eq. (\ref{BM}), the horizontal curve represents the global conservation
of energy.}
\label{fig:SS_cons}
\end{figure}

Three different families of exterior/initial scalar fields are showed in Table II. All them are specified in such a way that smooth metric functions across the matching surface $r=R$ are assured. For each set and any $R$, it is obtained $M$ as a function of the scalar field amplitude.
Taking as a reference the data set (a), for instance, it can be fitted the two others to it simply doing the appropriate rescaling in amplitude, as is showed figure \ref{fig:mass}. The limit mass for the background is $R/2$; this feature seems to be general. The limit mass to keep the incompressible static fluid is the well known Buchdhal mass limit $4R/9$. 
Therefore, if a black hole form it occurs by means of scalar radiation accretion.  

The exterior spacetime is determined integrating the field equations
(\ref{ekg+1}) and (\ref{ekg+2}).
To evolve the scalar wave on the specified background we have the set of ordinary differential
equations at $r=0$ for the scalar field (\ref{dotPhi0}) and (\ref{dotPhi1}).
This system requires only initial conditions for $\Phi_0$, $\Phi_1$ and $\Phi_2$, which are
fitted from the initial condition for the scalar field. Besides $M$ and $R$ are given as parameters, to define the matter distribution. 

The wave equation can be integrated following the null parallelogram method \cite{giw92}.
Note that,  as is required by the used method,
writing down (\ref{we1})
\begin{equation}
2g_{,ur}-[e^{2\beta}(1-2\tilde m/r)g_{,r}]_{,r} = -[e^{2\beta}(1-2\tilde m/r)]_{,r}g/r,
\label{we}
\end{equation}
where $g=r\Phi$, we have to take care considering the RHS. To avoid the numerical derivative,
that term is replaced by $-\Lambda g$, where 
\begin{equation}
\Lambda = \frac{e^{2\beta}}{r^2}\left\{ 4\pi r^2(p-\rho)+2\tilde m/r \right\}.
\label{ep}
\end{equation}

A subtle issue of this model is the Killing propagator to get
the energy conservation as showed in figure \ref{fig:balance}. 
Because the spacetime is fixed as a background, the right propagator is
\begin{equation}
\xi^\mu=e^{-2H}\delta^\mu_0.
\end{equation}
Figure \ref{fig:npc} displays the evolution of the initial data (a); it is evident that
the Newman--Penrose constant is conserved.
%%%%%%%%%%%%%%%%%%%%%%
\subsection{Scattering off a polytrope}
%%%%%%%%%%%%%%%%%%%%%%
In the case of a static polytrope 
\begin{equation}
p=K\rho^{\Gamma},
\end{equation} 
where $K$ is the polytropic constant and $\Gamma$ is the adiabatic exponent, related with the adiabatic index $n$ by $\Gamma=1+1/n$, the pressure and the energy density vanish at $r=R$. Again, we integrate numerically the system (\ref{e2}), (\ref{e3}) and (\ref{tov}) for any choice of $K$ and $\Gamma$, tipically $100$ and $2$ respectively. However, it is interesting to note that (\ref{tov}) written as
\begin{equation}
\frac{p_{,r}}{\rho + p} + \left[ \frac{1}{2}\ln(1-2\tilde m/r) + 2 \beta \right]_{,r} = 0,
\end{equation}
can be integrated to get
\begin{equation}
e^{2 \beta} = \frac{{F}^{1/2}(1-2\tilde m/r)^{-1/2}}{(1 + p/\rho)^{n+1}}.
\end{equation}
This result let us to conclude that no limit appears for the total mass distribution except the
black hole mass limit $R/2$. It clearly connects the surface with the center
of the distribution by means of $e^{2\beta_0}(1+p_0/\rho_0)^{n+1}=F^{1/2}$. 

In this model, a polytrope with a vacuum Schwarzschild exterior
as a background, the scalar field is scattered off and radiated to infinity. 
Figure \ref{fig:poly_cons} shows the energy conservation in this case. It should be stressed here that the scalar field gradient of the initial datum has to be zero at $r=R$ and the
scalar field itself partially immersed in the distribution.

\subsection{Quasinormal mode and late time tail decay}
For a Schwarzschild--Schwarzschild background an expected feature is showed when
an initial compact support scalar field (b) is evolved. In this case the distribution of matter corresponds to an incompressible fluid, that is, a Schwarzschild interior spacetime; the exterior is a Schwarzschild vacuum. Figure \ref{fig:qnm} display the quasinormal mode ringing and
the final tail decay. The energy conservation is displayed in figure \ref{fig:SS_cons}.

\section{Concluding remarks}
We have used corner stones in the characteristic formulation of general and numerical relativity to present in this paper a framework which couples matter with radiation \cite{bvm62}, \cite{b64}, \cite{tw66}. The old point of view of Bondian observers are mistakenly considered as Eulerian in numerical relativity for the mathematical treatment of matter \cite{b09}.

Although we only explored toy models, we believe they deserve future attention at least as initial conditions. The partially immersed scalar field in the fluid is hidden. This mixture constitutes a fixed and well behaved background to evolve a cloned scalar field. 
The face value of model A is its usefulness as a playground to study asymptotic and matching regions. It is striking how the matching leads us to a connection between the mass distribution and the amplitude
of the initial scalar field, and how the limit mass of $R/2$ does not depend apparently upon the initial data. The energy conservation and the Newman--Penrose constant tests give confidence
on the possible physical consequences of this simple model. The other two models represent simple
tests which are in agreement with expectations. 

We stress that this paper does not develop a general code, it just reports how Bondian observers
are plausible in Numerical Relativity. Numerics is simplified to show how these observers are
consistent with previous results. The underlaying problem, gravitational radiation coupled to matter, is in fact difficult. We show that even in the spherical case we have
to consider special regions, center and boundary surface carefully as well. We are currently
developing a general code based on section 3, which considers regularization and
matching regions in a clearer manner.

The most simple axisymmetric case from the Bondian point of view is out of scope of the
present investigation. There is a unique geometrical way to define Bondian observers in the absence of spherical symmetry: i) choose a tetrad to get the local Minkowskian frame; ii) make a double Lorentz boost to go to the frame comoving with the fluid. In this sense could be interesting to revise results from \cite{sfmp03}
to consider gravitational radiation considering explicitly Bondian frames and regularization
on special regions such as the axis of symmetry and boundary surface. 

An advanced study of incompressible--like fluids and polytropic matter indicates the relevance of central equation of state at the center of the star. This motivated us to prepare a hydrodynamic solver which will be reported elsewhere for an adiabatic situation.
A more general study of matter coupled to radiation within this framework is in progress.

\section*{Acknowledgments} 
We are grateful to Luis Rosales and Carlos Peralta for their comments. Finishing this work WB was on sabbatical leave from ULA. We thank to the referees for their valuable comments.

%\section*{References}

\end{document}